\documentclass[a4paper,11pt]{article}
\usepackage{pos}

\usepackage{placeins}
\usepackage{subcaption}

\usepackage{xspace}
\usepackage{natbib}

\renewcommand*{\thefootnote}{(\arabic{footnote})}

\newcommand{\pp}{\mathrm{pp}}

\newcommand{\sqrts}{\sqrt{s}}

\newcommand{\jpsi}{\mathrm{J}/\psi}

\newcommand{\pT}{p_\mathrm{T}}
\newcommand{\pTj}{p_{\mathrm{T},j}}
\newcommand{\pTmin}{p_{\mathrm{T,min}}}

\newcommand{\kT}{k_\mathrm{T}}
\newcommand{\HT}{\mathrm{H}_\mathrm{T}}

\newcommand{\pTij}{p_{\mathrm{T};i,j}}
\newcommand{\pTonetwo}{p_{\mathrm{T};1,2}}

\providecommand{\MGaMC}{{\sc\small MadGraph5\_aMC@NLO}}

\newcommand{\mgshort}{\textsc{MG5\_aMC}}
\newcommand{\pythia}{\textsc{Pythia}}
\newcommand{\pyeight}{\textsc{py8}}

\newcommand{\sigmaeffdps}{\sigma_\mathrm{eff,DPS}}
\newcommand{\sigmaefftps}{\sigma_\mathrm{eff,TPS}}

\newcommand{\sigmaeff}{\sigma_{\rm eff}}

\providecommand{\MGaMC}{{\sc\small MadGraph5\_aMC@NLO}}

\newcommand{\alpgen}{\textsc{Alpgen}}

\newcommand*{\eg}{e.g.,\@\xspace}
\newcommand*{\ie}{i.e.,\@\xspace}

\title{Six-jet production from triple parton scatterings in proton-proton collisions at the LHC}
\ShortTitle{Six-jet production from triple parton scatterings in pp collisions at the LHC}

\author*[a]{Marina Maneyro}
\author[b]{David~d'Enterria}
\affiliation[a]{Department of Physics, University of Liverpool,\\
L69 7ZE, Liverpool, United Kingdom}
\affiliation[b]{EP Department, CERN,\\
   CH-1211 Geneva, Switzerland}

\emailAdd{marina.maneyro-questa@liverpool.ac.uk}
\emailAdd{david.d'enterria@cern.ch}

\abstract{The production of six energetic jets in proton-proton ($\pp$) collisions at the LHC is studied as a means to directly observe for the first time the simultaneous scattering of three partons. The single-parton-scattering (SPS) cross sections for the production 2-, 3-, 4-, and 6-jets in $\pp$ collisions at center-of-mass energies of $\sqrts = 14$~TeV, are calculated up to next-to-leading-order (NLO) accuracy in perturbative quantum chromodynamics with the \MGaMC\ and \alpgen\ codes complemented with \pythia-8 for parton showering, hadronization, and decays. Jets are reconstructed using the anti-$\kT$ algorithm with distance parameter $R=0.4$. Assuming factorization of multiple hard-scattering probabilities in terms of SPS cross sections, the contributions to six-jet production from double- (DPS) and triple- (TPS) parton scatterings are derived. We find that the TPS contributions represent a ${\approx}20\%$ (${\approx}1\%$) fraction of the total 6-jets yields for minimum jet transverse momenta of $\pTmin = 20$~(40)~GeV. A detailed multivariate analysis with realistic simulations of fully reconstructed jet samples for the TPS signal and DPS and SPS backgrounds indicates that TPS can be observed in events with six jets with $\pTmin=40$~GeV each, by collecting an integrated luminosity of $\mathcal{O}(50$~pb$^{-1}$) in a dedicated low-pileup run at the LHC.}

\FullConference{31st International Workshop on Deep Inelastic Scattering (DIS2024)\\
 8–12 April 2024\\
Grenoble, France\\}

\begin{document}
\maketitle

\section{Introduction}

Hadronic collisions at high energies are characterized by multiple simultaneous collisions of their underlying partonic (quark and gluon) degrees of freedom~\cite{Bartalini:2018qje}. At the multi-TeV collision energies of the CERN Large Hadron Collider (LHC), the study of processes leading to the production of two heavy particles -- such as quarkonia~\cite{Kom:2011bd,Baranov:2012re,Lansberg:2014swa,Chapon:2020heu} and/or electroweak bosons~\cite{Mekhfi:1983az,Kulesza:1999zh,Maina:2009sj,Gaunt:2010pi} -- and/or pairs of energetic jets~\cite{Paver:1982yp, Humpert:1984ay,Berger:2009cm,Blok:2010ge,Fedkevych:2020cmd} in the same proton-proton ($\pp$) collision, known as double-parton scatterings (DPS), has attracted increasing experimental and theoretical attention as a means to study the generalized parton densities~\cite{Diehl:2011yj}, the transverse parton profile of the protons (in particular, its evolution with collision energy)~\cite{Rinaldi:2018bsf}, and the correlations in kinematics and quantum numbers among partons inside the hadronic wavefunctions~\cite{Calucci:2010wg,Blok:2017alw}. 
More recently, it has been also proposed to investigate triple-parton scatterings (TPS) in $\pp$ and proton-nucleus collisions~\cite{dEnterria:2016ids, dEnterria:2016yhy} to further improve our understanding of multiple hard partonic scatterings~\cite{dEnterria:2017yhd}. Studies of TPS have thereby been proposed in final states with triple charm and triple bottom~\cite{dEnterria:2016ids, dEnterria:2016yhy}, triple $\jpsi$ mesons~\cite{Shao:2019qob}, and Z boson plus four jets~\cite{Andersen:2023hzm}. 
The study of multijet (in particular, four-jet) production has been successfully exploited in the past to observe and study DPS processes at colliders~\cite{AxialFieldSpectrometer:1986dfj,UA2:1991apc,CDF:1993sbj,CDF:1997lmq,CDF:1997yfa,CMS:2016kdy, ATLAS:2016rnd,CMS:2021lxi}. The purpose of this work, reported in more detail in~\cite{DdEManeyro}, is to consider the case of six-jet production in $\pp$ collisions at the LHC as a means to observe TPS processes for the first time.

The simplest phenomenological description of DPS is based on a simple geometric model of the partonic transverse profile of the colliding hadrons disregarding any parton correlations. In such a picture, the probability to produce $n$ high-$\pT$ and/or heavy particles is proportional to the product of the probabilities of producing them independently in single parton scatterings (SPS). For the DPS case, the cross section to produce two particles $X_1$ and $X_2$ can be then determined via
\begin{equation}
\sigma^{\pp \to X_1\,X_2}_\mathrm{DPS} = \left(\frac{m}{2}\right)\, \frac{\sigma^{\pp \to X_1}_\mathrm{SPS} \, \sigma^{\pp\to X_2}_\mathrm{SPS}}{\sigmaeff},
\label{eq:pocketDPS}
\end{equation}
where $\sigma^{\pp \to X_{1,2}}_\mathrm{SPS}$ are the inclusive SPS cross sections for the production of $X_1$ and $X_2$, which are experimentally measurable and/or calculable with perturbative Quantum Chromodynamics (pQCD) methods; the integer $m$ is a combinatorial factor to avoid double counting the same process: $m=1\,(2)$ if $X_1=X_2$ ($X_1\neq X_2$); and the effective cross section $\sigmaeff$ in the denominator ensures the appropriate dimensionality of the DPS cross section. In the purely geometric approach, the value of $\sigmaeff$ can be estimated from the integral of the transverse overlap function $T(\mathbf{b})$ over impact parameter $b$ in the $\pp$ collision, where $T(\mathbf{b})$ depends on the transverse parton density of the proton $\rho(\mathbf{b})$~\cite{dEnterria:2020dwq}. The wealth of experimental DPS studies at the LHC and Tevatron indicate a value of the effective cross section around $\sigmaeff\approx 15$~mb~\cite{Bartalini:2018qje} used hereafter for numerical estimates. Under the assumption of factorization of hard scattering contributions and absence of correlations, the TPS cross section for the production of three processes ($X_1$, $X_2$, $X_3$), can be similarly written as~\cite{dEnterria:2016ids}:
\begin{equation}
\sigma^{\pp \to X_1\,X_2\,X_3}_\mathrm{TPS} = \left(\frac{m}{3!}\right)\, \frac{\sigma^{\pp \to X_1}_\mathrm{SPS} \, \sigma^{\pp\to X_2}_\mathrm{SPS}\, \sigma^{\pp\to X_3}_\mathrm{SPS}}{\sigmaefftps^2}
\; \approx\; \left(\frac{m}{4}\right)\, \frac{\sigma^{\pp \to X_1}_\mathrm{SPS} \, \sigma^{\pp\to X_2}_\mathrm{SPS}\, \sigma^{\pp\to X_3}_\mathrm{SPS}}{\sigmaeff^2}
\,,
\label{eq:pocketTPS}
\end{equation}
\ie\ as the product of the corresponding SPS cross sections normalized by a combinatorial factor ($m = 1$ if $X_1=X_2=X_3$; $m = 3$ if $X_i=X_j\neq X_k$ for $(i,j,k)=\{1,2,3\}$; and $m = 6$ if $X_1\neq X_2\neq X_3$) and TPS effective cross section, $\sigmaefftps$. The latter appears squared to preserve the proper units of the result, and its numerical value is closely related to its DPS counterpart $\sigmaeff$. In the geometric picture, both effective cross sections are related by $\sigmaefftps = \kappa\cdot\sigmaeff$ with proportionality constant $\kappa = 0.82\pm 0.11$~\cite{dEnterria:2016ids}, yielding the second (approximate) equality of Eq.~(\ref{eq:pocketTPS}).

\begin{figure}[htpb!]
    \centering
\includegraphics[width=.99\textwidth]{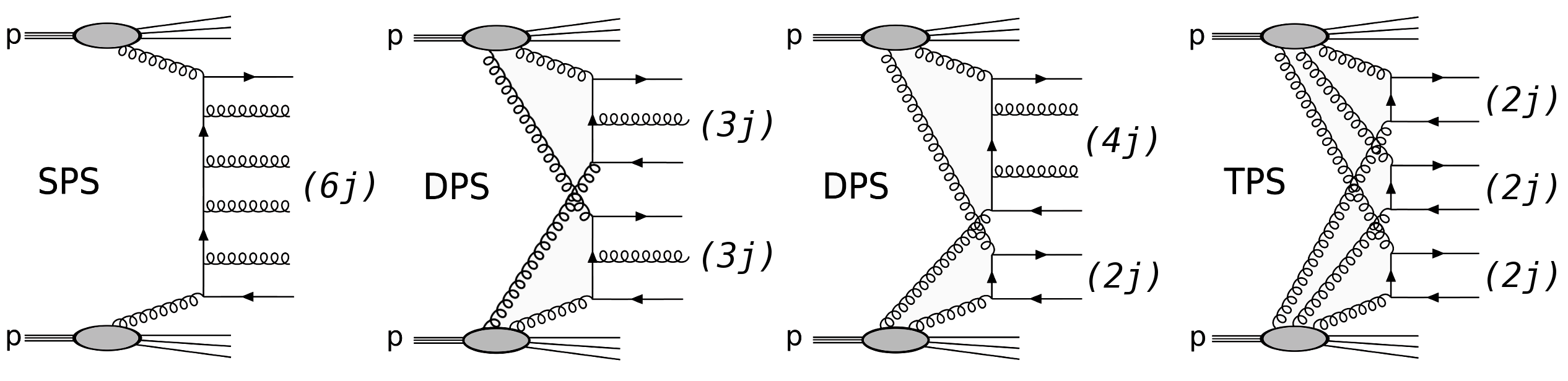}
\caption{Representative diagrams for the production of 6-jets in SPS (leftmost), DPS (center, with DPS1 ($2j+4j)$) and DPS2 ($3j+3j$), and TPS (rightmost) processes in $\pp$ collisions.}
\label{fig:6jets_TPS}
\end{figure}

The TPS process of interest in this work is that of 6-jet production via $\pp\to2j+2j+2j$ shown in the rightmost diagram of Fig.~\ref{fig:6jets_TPS}. The production of the same final state proceeds, however, with higher probability through the SPS ($\pp\to 6j$) and DPS ($\pp\to 3j+3j$ and $\pp\to2j+4j$) processes displayed, respectively, in the leftmost and central diagrams of Fig.~\ref{fig:6jets_TPS}.

\vspace{-0.5cm}

\FloatBarrier
\section{Results}

We first compute the expected theoretical cross sections for all processes contributing to 6-jet production, via the diagrams of Fig.~\ref{fig:6jets_TPS}, above a given minimum $\pTmin$ value. 
The SPS results have been obtained at next-to-leading-order (NLO) accuracy with \MGaMC\ (\mgshort\ hereafter)~\cite{Alwall:2014hca} whenever possible, and at LO with \alpgen~\cite{Mangano:2002ea} for the highest jet multiplicities ($4j$ and $6j$) for which the NLO calculations are too time-consuming or only available in analytical form~\cite{Bern:2011ep,Badger:2012pf}. The NNPDF4.0 NLO and LO parton distribution functions (PDFs)~\cite{NNPDF:2021njg} are used. The default renormalization and factorization scales are set to half the sum of scalar transverse momenta of the jets in the event $\mu_\mathrm{R}=\mu_\mathrm{F} = \HT/2$, with  $\HT=\sum^{N_\text{partons}}_i p_{\mathrm{T},i}$, and $\mu_\mathrm{R,F}$ are independently varied within a factor of two up and down to assess the theoretical uncertainties from missing higher order corrections. The generated partons are showered and hadronized with \pythia~8 (\pyeight\ hereafter)~\cite{Sjostrand:2014zea}, and the jets are reconstructed with the anti-$\kT$ algorithm~\cite{Cacciari:2008gp}, implemented in \textsc{FastJet}~\cite{Cacciari:2011ma}, with distance parameter $R=0.4$ (for \alpgen, we also impose a minimum jet separation of $\Delta R>0.8$).
The NLO results are obtained with a small asymmetry ($\delta\pT \approx 5$~GeV) around the threshold $\pTmin$ so that there is phase space allowed for additional gluon radiation between the leading and subleading jets~\cite{Frederix:2016ost}. From the SPS $2j,\,3j,\,4j, 6j$ production cross sections calculated for $\pp\,$(14\,TeV) collisions within the typical rapidity coverage of the ATLAS and CMS experiments ($|\eta^j|<5$), we estimate the corresponding DPS and TPS cross sections with Eqs.~(\ref{eq:pocketDPS}) and (\ref{eq:pocketTPS}) using $\sigmaeff = 15$~mb.

\begin{figure*}[htpb!]
\centering
\includegraphics[width=0.53\textwidth]{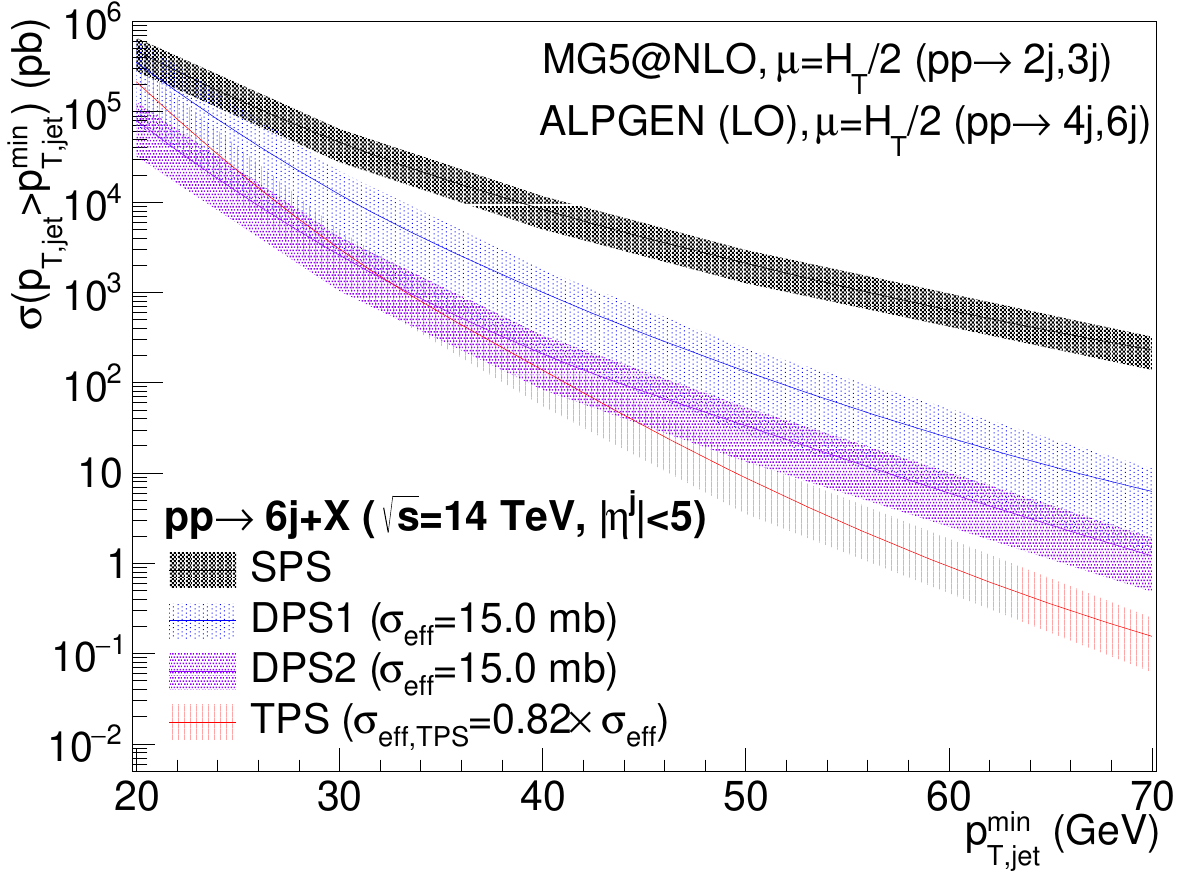}
\includegraphics[width=0.46\textwidth]{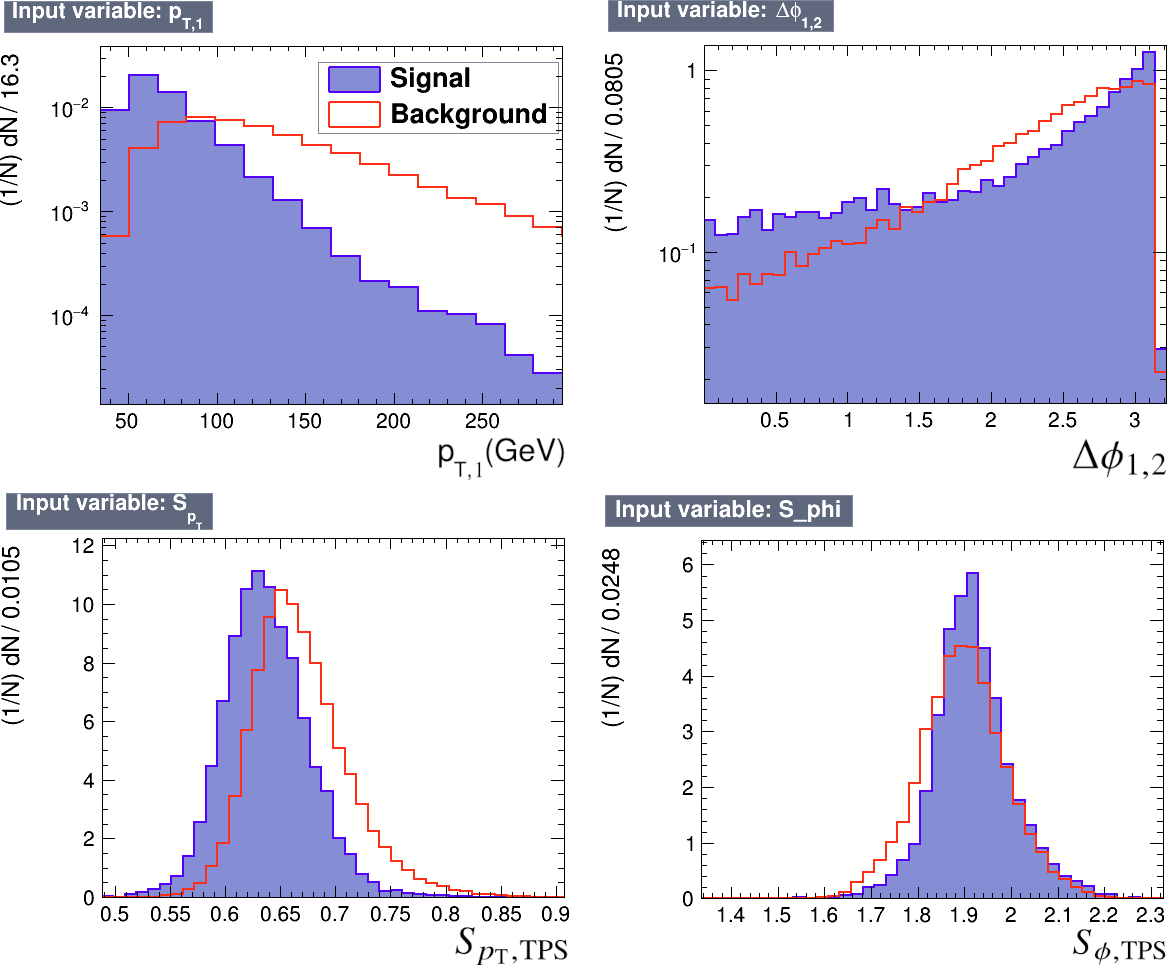}

\caption{Left: Integrated cross sections for the production of 6 jets above a given $\pTmin$ in pp\,(14~TeV) from SPS, DPS, and TPS processes (for $\sigmaeffdps=15$~mb). Right: Examples of normalized distributions for BDT variables to discriminate the TPS signal (blue filled histogram) and SPS\,+\,DPS backgrounds (red histogram) in the $\pp\to 6j$ analysis.
\label{fig:SPS,DPS,TPS_frac}}
\end{figure*}

The predicted SPS, DPS, and TPS 6-jets cross sections are shown as a function of $\pTmin$ in Fig.~\ref{fig:SPS,DPS,TPS_frac} (left). Within the relatively large uncertainties, the SPS contributions dominate the cross sections, followed by the DPS1 ($2j+4j$), DPS2 ($3j+3j$) and TPS components. The TPS contributions decrease more rapidly with $\pTmin$ due to the cube dependence on the SPS cross section. 
The DPS and TPS contributions account for about 20\% of the 6-jet yields at $\pTmin \approx 20$~GeV, a value which is however too low for precisely reconstructing them experimentally (because of large nonperturbative corrections, jet energy scale and resolution effects, and pileup). We therefore chose as working point a $\pTmin = 40$~GeV threshold for the 6-jets study, where TPS represent about 1\% of the total yields.
The corresponding cross sections are listed in Table~\ref{tab:pp_xsections}. The LO results obtained with \mgshort$\,+\,$\pyeight\ and \alpgen$\,+\,$\pyeight\ are consistent with each other for all jet multiplicities, and have ${\approx}30\%$ theoretical uncertainties (few percent PDF uncertainties are neglected). The NLO predictions basically match the LO results, but with reduced scale uncertainty. More precise and accurate cross sections can be achieved with calculations with a higher degree of pQCD accuracy~\cite{Currie:2017eqf,Chen:2022tpk,Czakon:2021mjy}, and/or by directly using the experimental data, for the individual $2j,\,3j,\,4j$ categories.

\begin{table}[htpb!]
\centering
\caption{Cross sections for 2-, 3-, 4-, 6-jet SPS production ($R=0.4$ anti-$\kT$ jets) with $\pTj> 40$~GeV~and $|\eta_j|<5$ in $\pp$ collisions at $\sqrts = 14$~TeV computed at LO and NLO accuracy with  \alpgen$\,+\,$\pyeight\ and \mgshort$\,+\,$\pyeight, and corresponding DPS1 ($2j+4j$), DPS2 ($3j+3j$), and TPS ($2j+2j+2j$) cross sections derived from them via Eq.~(\ref{eq:pocketDPS})--(\ref{eq:pocketTPS}). The quoted uncertainties (\%) are mostly due to theoretical scale variations.}
\label{tab:pp_xsections}
\vspace{-0.4cm}
\resizebox{\textwidth}{!}{%
\begin{tabular}{l|cccc} \hline
Process  & \alpgen\ (LO) & \mgshort (LO) & \mgshort (NLO) \\\hline
$\pp\to 2j$ (SPS) & 48~$\mu$b & 47~$\mu$b $\substack{+30\% \\ -20\%}$ &  50~$\mu$b $\substack{+15\% \\ -10\%}$ \\

$\pp\to 3j$ (SPS) & 2.7~$\mu$b  & 2.5~$\mu$b $\substack{+40\% \\ -30\%}$ &  2.5~$\mu$b  $\substack{+20\% \\ -15\%}$ \\
$\pp\to 4j$ (SPS) & 0.4~$\mu$b & 0.3~$\mu$b $\substack{+40\% \\ -30\%}$ &---  \\
$\pp\to 6j$ (SPS) & $8.2~\substack{+40\% \\ -30\%}$~nb   & --- & --- \\\hline
$\pp\to 6j$ (DPS1), Eq.~(\ref{eq:pocketDPS}) with $\sigmaeff = 15$~mb &  &  & $1.0~\substack{+45\% \\ -30\%}$~nb\\ 
$\pp\to 6j$ (DPS2), Eq.~(\ref{eq:pocketDPS}) with $\sigmaeff = 15$~mb &  &  & $200~\substack{+80\% \\ -60\%}$~pb \\ 
$\pp\to 6j$ (TPS), Eq.~(\ref{eq:pocketTPS}) with $\sigmaeff = 15$~mb  &  &  & $140~\substack{+40\% \\ -30\%}$~pb\\\hline
\end{tabular}
}
\end{table}

In order to identify the TPS contributions to the total 6-jet yields, we exploit their expected kinematic differences with respect to DPS and SPS events. Six jets from TPS should mostly appear as three pairs of jets back-to-back each in azimuth and balanced in $\pT$, whereas the leading jets from SPS contributions should have harder $\pT$ values and subleading jets should lack significant pair-wise correlations. In order to identify the TPS signal, multiple kinematic variables are defined and introduced in a multivariate analysis (MVA) based on realistic parton-showered and hadronized simulations of 6-jets events generated with \mgshort+\pyeight\ and \alpgen+\pyeight. 

The 6-jet events samples are obtained by appropriately merging the generated SPS $2j, 3j, 4j$ events weighted according to Eqs.~(\ref{eq:pocketDPS}) and (\ref{eq:pocketTPS}). The resulting $2j+4j$, $3j+3j$ and $2j+2j+2j$ data samples, alongside with the $6j$ SPS background events, are subsequently analyzed requiring that at least 6 reconstructed jets pass the $(\pT,\eta)$ acceptance cuts. Such a procedure also accounts for any efficiency loss due to, \eg\ potential jet overlaps in the event merging. The reconstruction and acceptance efficiencies for SPS, DPS, and TPS 6-jets are about 50\%, 50--30\%, and 30\%, respectively, driven mostly by the softer nature of the TPS jets. Next, we define more than 50 different single-jet and multijet variables that are fed to a MVA based on boosted decision trees (BDTs) implemented in the TMVA framework~\cite{TMVA:2007ngy}. 
The first classifier variable of interest is the $\pTj$ of each of the 6 jets, used to sort them decreasingly so that $j=1$ always refers to the leading jet of any given event. We define also variables for each possible jet pair combination, including invariant masses $M_{i,j}$, pair's balance in $\pT$ and separation in $\eta$ and $\phi$: $\Delta{\pTij}=|\Vec{p}_{\mathrm{T},i}+\Vec{p}_{\mathrm{T},j}|/|\Vec{p}_{\mathrm{T},i}|+|\Vec{p}_{\mathrm{T},j}|,\;  \Delta\eta_{i,j}=|\eta_i-\eta_j|, \;  \Delta\phi_{i,j}=|\phi_i-\phi_j|$, for $i,j=1,...,6$. More advanced observables are defined based on generalizations of variables suggested in previous 4-jets DPS studies~\cite{UA2:1991apc,Berger:2009cm}, such as
\begin{equation}
S_{\pT,\rm TPS}^2\equiv\frac{1}{3}\left[\left(\cfrac{|\Vec{p}_{\mathrm{T},i}+\Vec{p}_{\mathrm{T},j}|}{|\Vec{p}_{\mathrm{T},i}|+|\Vec{p}_{\mathrm{T},j}|}\right)^2+\left(\cfrac{|\Vec{p}_{\mathrm{T},k}+\Vec{p}_{\mathrm{T},l}|}{|\Vec{p}_{\mathrm{T},k}|+|\Vec{p}_{\mathrm{T},l}|}\right)^2+\left(\cfrac{|\Vec{p}_{\mathrm{T},m}+\Vec{p}_{\mathrm{T},n}|}{|\Vec{p}_{\mathrm{T},m}|+|\Vec{p}_{\mathrm{T},n}|}\right)^2\right]\,,
\label{eq:s}
\end{equation}
which is sensitive to the expected $\pT$ balance correlations among jet pairs corresponding to the independent interactions of three constituent partons in TPS production.
A similar variable in structure to $S_{\pT}$ focuses instead on the $\phi$ correlations among jet pairs:
\begin{equation}
S_{\phi,\rm TPS}^2\equiv\frac{1}{3}\left[|\phi_i-\phi_j|^2+|\phi_k-\phi_l|^2+|\phi_m-\phi_n|^2\right].
\label{eq:sphi}
\end{equation}
Additional $S_{\pT,\rm DPS}$ and $S_{\phi,\rm DPS}$-type variables are defined involving only two jet pairs at a time to ensure good discrimination also between DPS and TPS events.
The output files with all variables from the SPS, DPS, and TPS samples are fed into the TMVA for training and testing purposes. Examples of distributions of BDT variables for signal (blue solid histogram) and background (red histogram) are shown in Fig.~\ref{fig:SPS,DPS,TPS_frac} (right). One can see that the leading jet $\pT$ from the TPS sample has a softer distribution than that of the SPS+DPS backgrounds. The top discriminating variables identified by the MVA are: $p_{\mathrm{T},1}$, $S_{\phi,\rm TPS}$, $S_{\phi,\rm DPS}$, $\Delta\phi_{1,2}$, $\Delta_{\pTonetwo}$, and $S_{\pT,\rm TPS}$.

Analysis of the BDT response allows the determination of the number of expected TPS signal ($\mathcal{S}$) and associated DPS and SPS background events ($\mathcal{B}$) that are needed to reach a statistical significance of 5 standard deviations. We found that, for the assumed $\sigmaeff = 15$~mb effective cross section and threshold jet $\pTmin = 40$~GeV, an integrated luminosity of 50~pb$^{-1}$, corresponding to $\mathcal{S} \approx 2300$ (signal) and $\mathcal{B}\approx 2\cdot 10^5+3\cdot10^4$ (SPS\,+\,DPS background) events, respectively, would allow to achieve experimental observation of TPS 6-jet production above the background-only hypothesis. Such a relatively small amount of data should be, however, recorded under low-pileup conditions so as to precisely reconstruct jets with $\pTmin \approx 40$~GeV, and eliminate the possibility of different pp multijet events contaminating the 6-jets event sample. Further details on the analysis and the final results will appear in Ref.~\cite{DdEManeyro}.

\paragraph*{Acknowledgments.---} Support from the European Union's Horizon 2020 research and innovation program (grant agreement No.\ 824093, STRONG-2020) is acknowledged.

\bibliographystyle{JHEP}
\bibliography{biblio}

\end{document}